\documentclass[11pt]{article}
\usepackage{amsfonts}
\usepackage{amsmath}
\usepackage{amscd}

\usepackage{epsf}
\usepackage{times}
\usepackage{latexsym}
\usepackage{graphics}
\usepackage{graphicx}

\begin{document}
\title{\bf  Factorizations of one
dimensional \\ classical systems}

\author{ \c{S}eng\"ul Kuru$^{\dag \ddag}$ and Javier Negro$^{\dag *}$}

\maketitle

\begin{center}
{\it $\dag$ \ Departamento de F\'{\i}sica Te\'{o}rica, At\'omica y \'Optica,\\
Universidad de Valladolid, 47071 Valladolid, Spain}

{\it $\ddag$\ Department of Physics, Faculty of Science, \\Ankara University,
06100 Ankara, Turkey}
\end{center}

\begin{abstract}
A class of one dimensional classical systems is characterized from
an algebraic point of view. The Hamiltonians of these systems are
factorized in terms of two functions that together with the
Hamiltonian itself close a Poisson algebra. These two functions lead
directly to two time-dependent integrals of motion from which the
phase motions are derived algebraically. The systems so obtained
constitute the classical analogues of the well known factorizable
one dimensional quantum mechanical systems.
\end{abstract}

\vskip1cm


\bigskip


\bigskip

\noindent
$^*$ {\it Corresponding author}. Phone: (34)-983423040. FAX:
(34)-983423754. E-mail: jnegro@fta.uva.es

\vfill\eject

\section{Introduction}

The one dimensional systems, with a time independent Hamiltonian,
are treated in a rather different way in classical and quantum
mechanics. In quantum mechanics the Schr\"odinger equation for such
systems determines the spectrum that can include a discrete part for
bound states and a continuous sector for unbounded (scattering) states. For periodic potentials there are forbidden and
allowed bands of energy whose borders constitute the spectrum in a
wider sense. Only in some cases the eigenvalue problem for the
stationary Schr\"odinger equation can be solved ``exactly'' so that
we are able to find a ``closed expression'' for the spectrum (energy
eigenvalues) as well as for the corresponding eigenfunctions. This
situation arises (but it is not the only option) when there are
lowering and raising operators that close a spectrum generating
algebra with the Hamiltonian and factorize the Casimir of the
algebra up to a constant \cite{Infeld,cooper,raab,lemus,negro}.

From the point of view of classical mechanics, the Hamiltonians of
one dimensional conservative systems are integrals of motion, hence
all these systems are maximally integrable \cite{evans}. This
property, in classical mechanics, implies that every closed and
bounded trajectory must be periodic. For this reason, often they are
considered almost trivial and usually have not received much
attention \cite{perelomov}. The values of the energy are continuous
for both, bound and unbounded motion states. While the motion is
necessarily periodic with a certain frequency (in general depending on the energy) for bound states, it is non-periodic for the unbounded
ones. In some special cases, 
we can find the explicit expression for the motion in the phase space by means of the algebraic structure underlying these classical systems.
The main purpose of this paper is to show how this structure is
presented for a class of one dimensional classical systems. The
Hamiltonians that allow for this treatment are classical analogues of
some quantum systems. We will see that the algebraic structure of
these quantum and classical systems are similar, but at the same
time there are important differences. Therefore, the general
motivation of this work is to close the symmetry considerations of
both, classical and quantum mechanics in this framework.

The organization of the paper is as follows. In section 2 we
introduce our algebraic approach to the classical systems in terms
of Poisson algebras \cite{bonatsos,daskaloyannis,daskaloyannis2}.
For the factorizable classical systems we find two time-dependent
integrals of motion similar to the Bohlin invariants
\cite{gettys,ray,pilar}. These invariants allow us to obtain
algebraically the solutions of motion that give rise to trajectories in the phase space. Usually, the time-independent integrals of motion have been used to simplify the
differential equation of motion in order to get explicit solutions,
but the time-dependent integrals of motion have been hardly used in
studying stationary systems. This work is one step forward in the
applications of such a type of invariants.
This development is systematically
applied to some examples in sections 3, 4 and 5. We will end the paper
with some comments and conclusions on the results here obtained.

\section{Factorizations and deformed algebras}

Let us consider the Hamiltonian
\begin{equation}\label{h}
H(x,p)=\frac{p^2}{2\,m}+V(x)
\end{equation}
where $x$, $p$  are canonical coordinates, i.e., $\{p,x\}=1$, being
$\{\cdot,\cdot\}$ the notation for the Poisson brackets, and $V(x)$
is the potential. Hereafter we will set $2\,m=1$ for simplicity. We
will investigate a kind of factorization of the Hamiltonian $H$ in
terms of two complex-conjugate functions $A^{\pm}$ as
\begin{equation}\label{hh}
H=A^{+}A^{-}+\gamma(H)\ .
\end{equation}
Here the term $\gamma(H)$, contrary to the usual factorizations in
quantum mechanics, may depend on $H$. In the case of bound states
with positive energy ($H>0$), we will
consider $A^{\pm}$ linear in $p$ and having the form, as suggested
from (\ref{hh}),
\begin{equation}\label{a}
A^{\pm}=\mp i f(x)\,p+\sqrt{H}\,g(x)+\varphi(x)+\phi(H)\, 
\end{equation}
where the functions $f(x)$, $g(x)$, $\varphi(x)$ and $\phi(H)$ will
be determined in each case.
It is clear that a global constant factor in the functions $A^\pm$
will produce an equivalent factorization so that in the following we
will choose this factor to get simpler expressions.

We will ask the functions, $A^{\pm}$ and $H$, to define a deformed
algebra with the Poisson brackets \cite{bonatsos,daskaloyannis} as
follows
\begin{equation}\label{ha}
\{H,A^{\pm}\}=\pm i\,\alpha(H)\,A^{\pm}
\end{equation}
\begin{equation}\label{aa}
\{A^{+},A^{-}\}=-i \,\beta(H).
\end{equation}
The auxiliary functions $\alpha(H)$, $\beta(H)$, and $\phi(H)$ will
be expressed in terms of the powers of $\sqrt{H}$. In the case of
bound motions with negative energy we must replace the square roots 
$\sqrt{H}$ by $\sqrt{-H}$ in the above construction. There are
systems that can allow for bound and unbounded motions, in such cases
the complex character of the factors $A^{\pm}$ can change in each
energy sector, as well as the deformed algebra. This change also
happens in the quantum frame for the algebras describing bound and
scattering states of the same system \cite{alhassid}.

Relation (\ref{ha}) is the most important since it implies both
(\ref{hh}) and (\ref{aa}). In order to show this, we compute
\begin{equation}\label{haa}
\{H,A^{+}A^{-}\}=A^{+}\{H,A^{-}\}+\{H,A^{+}\}A^{-}=0
\end{equation}
where we have made use of  (\ref{ha}). From the vanishing of this
bracket we conclude that $A^+A^-$ will depend only on $H$ and
therefore, we can express this dependence as in (\ref{hh}). To show
(\ref{aa}), we start from the Jacobi identity
\begin{equation}\label{ji}
\{H,\{A^{+},A^{-}\}\}=-\{A^{+},\{A^{-},H\}\}-\{A^{-},\{H,A^{+}\}\}
\end{equation}
substituting on the r.h.s of this identity relation (\ref{ha}), we
find
\[
\{H,\{A^{+},A^{-}\}\}= -i\{ A^+A^-,\alpha(H)\}=0
\]
so that the bracket $\{A^{+},A^{-}\}$ can be expressed in the
form (\ref{aa}).

Now, for a system allowing for this kind of factorization, we can
construct two time-dependent integrals of motion,
\begin{equation}\label{q}
Q^{\pm}=A^{\pm}e^{\mp\,i\,\alpha(H)\,t}.
\end{equation}
It is easy to check that the total time derivative of $Q^\pm$ is equal to zero:
\begin{equation}\label{td}
\frac{d Q^{\pm}}{dt}=\{H,Q^{\pm}\}+\frac{\partial Q^{\pm}}{\partial
t}=0
\end{equation}
since the nonzero partial time derivative is
compensated by the poisson bracket of the integral of motion with the
Hamiltonian by using relation (\ref{ha}).
The (eigen)values of the integrals of motion $Q^{\pm}$
will be denoted
\begin{equation}\label{q2}
q^\pm=c\, e^{\pm i\,\theta_0}
\end{equation}
where $c = |q^\pm|$. Notice that this expression can also change for unbounded
motions. Having in mind the factorization (\ref{hh}), the modulus
$c$ of $q^\pm$ will depend on the (eigen)value $E$ of the Hamiltonian
$H$: $c= c(E)$.

The two independent integrals $Q^{\pm}(x,p,t)=c(E)\, e^{\pm i\,\theta_0}$ allow us to find algebraically
the trajectories ($x(t),\, p(t)$) in the phase space. In particular,
we can appreciate from (\ref{q}) that the frequencies of the motion
for bound states are given by $\alpha(E)$.

If we substitute $p^2=H-V(x)$ and $A^\pm$ given by (\ref{a}) in (\ref{ha}) we get two
relations for the unknown functions $g(x)$,  $f(x)$, $\alpha(H)$, $\phi(H)$,
$\varphi(x)$ and $V(x)$:
\begin{equation}\label{fg}
\begin{array}{ll}
f(x)=\frac{2}{\alpha(H)}\left(\sqrt{H}g'(x)+\varphi'(x)\right)
\\[2.ex]
-2f'(x)(H-V(x))+f(x)\,V'(x)=\alpha(H)\left(\sqrt{H}\,g(x)+\phi(H)+\varphi(x)\right)
\end{array}
\end{equation}
where the prime denotes the derivative with respect to the
corresponding argument of each function. If we use $A^{\pm}$  in (\ref{hh}), we have the following equation for  $\gamma(H)$
\begin{equation}\label{hf}
H=f^2(x)(H-V(x))+\left(\sqrt{H}\,g(x)+\phi(H)+\varphi(x)\right)^2+\gamma(H)
\end{equation}
which will determine completely the potential. Finally, substituting $A^{\pm}$
into (\ref{aa}) we will get an equation for the remaining function
$\beta(H)$,
\begin{equation}\label{b}
\begin{array}{l}
 \beta(H)=\sqrt{H}\,\big[2
\,f(x)\,g'(x)-2\,f'(x)\,g(x)\big]+\frac{1}{\sqrt{H}}\big[f(x)\,g(x)\,V'(x)+2\,f'(x)\,g(x)\,V(x)
\big]\\[2.ex]
\qquad
-4\,f'(x)\,\phi'(H)(H-V(x))+2\,f(x)\,\big[V'(x)\,\phi'(H)+\varphi'(x)\big].
\end{array}
\end{equation}

In the next sections we will study systematically some examples. To
get the potentials and the functions  $f(x)$, $g(x)$, $\varphi(x)$,
$\alpha(H)$, $\gamma(H)$ and $\beta(H)$, first we have chosen a
suitable  form of  $\phi(H)$, $g(x)$ and $\varphi(x)$, and we have also assumed that $H$ may have a positive or a negative character.

\section{Simple potentials:
oscillator, Scarf and P\"oschl-Teller}

In this section, we will consider the simplest solutions of the
problem which are obtained when the last term, $\phi(H)$, in the
expression (\ref{a}) of $A^\pm$ vanishes.
Now, from relation (\ref{hf}) we see that there are two kind
of solutions: those with $(g(x)=0, \varphi(x)\neq 0)$ and
the complementary set $(g(x)\neq 0, \varphi(x)= 0)$.
The first choice leads to the harmonic oscillator, and the second
one to the Scarf and P\"oschl-Teller potentials as shown below.

\subsection{The harmonic oscillator $(g=0, \varphi\neq0)$}

This choice gives rise to
\begin{equation}\label{aho}
A^{\pm}=\mp i f(x)\,p+\varphi(x)
\end{equation}
and using it in (\ref{fg}), we have
\begin{equation}\label{ffa}
f(x)=\frac{2}{\alpha(H)}\,\varphi'(x)
\end{equation}
\begin{equation}\label{fhva}
-2\,f'(x)\,(H-V(x))+f(x)\,V'(x)=\alpha(H)\,\varphi'(x).
\end{equation}
Here, since we assume that $H$ is a variable independent of $x$, then Eq. (\ref{ffa}) gives us $\alpha(H)=\alpha_{0}$ and
(\ref{ffa})-(\ref{fhva}) take the form
\begin{equation}\label{ffa1}
f(x)=\frac{2}{\alpha_{0}}\,\varphi'(x)
\end{equation}
\begin{equation}\label{fhva1}
-2\,f'(x)\,H+2\,f'(x)\,V(x)+f(x)\,V'(x)=\alpha_{0}\,\varphi'(x).
\end{equation}
Taking into account that the coefficient of $H$  in (\ref{fhva1}) has to be equal to zero,
since it is the only term depending on $H$, from (\ref{ffa1}) we have $\varphi(x)=a_0\,x+b_{0}$. The
remanning part of (\ref{fhva1}) gives us
\begin{equation}\label{fva}
V'(x)=\frac{\alpha_{0}^{2}\,\varphi(x)}{2\,\varphi'(x)}.
\end{equation}
Then, substituting $\varphi(x)$ in this equation and
integrating, we get the harmonic oscillator potential
\begin{equation}\label{fva1}
V(x)=\frac{\alpha_{0}^2}{2\,a_0}\left(\frac{a_0\,x^2}{2}+b_{0}\,x\right)+c
\end{equation}
where $c$ is integration constant. From Eq.~(\ref{hf}), we see that
$a_0={\alpha_{0}}/{2}$, and $\gamma_{0}=c-b_{0}^2$, so the potential
(\ref{fva1}) becomes
\begin{equation}\label{v1}
V(x)=\left(\frac{\alpha_{0}}{2}x+b_{0}\right)^2+\gamma_{0}.
\end{equation}
Using (\ref{b}), we get $\beta(H)=\alpha_{0}$ and therefore, in this case we can
rewrite all the functions as
\begin{equation}\label{afp1}
\begin{array}{lll}
f(x)=
1,\quad & g(x) = 0,\quad &\varphi(x)=\frac{\alpha_{0}}{2}x+b_{0}
\\[2.ex]
\alpha(H)=\alpha_{0},\quad & \beta(H)=\alpha_{0},\quad &
\gamma(H)=\gamma_{0}.
\end{array}
\end{equation}
The explicit expressions of the factor functions are
\begin{equation}\label{a1}
A^{\pm}=\mp i\,p+(\frac{\alpha_{0}}{2}x+b_{0})
\end{equation}
and the Hamiltonian  is factorized in terms of these functions as
\begin{equation}\label{haab}
H = A^+A^- +\gamma_0.
\end{equation}
Finally, the Poisson brackets read
\begin{equation}\label{ha1}
\{H,A^{\pm}\}=\pm i\,\alpha_{0}\,A^{\pm},\qquad \{A^{+},A^{-}\}=-
i\,\alpha_{0}.
\end{equation}
Of course, they constitute the oscillator algebra realized in terms of
Poisson brackets. Now, from (\ref{q}) we can write the
time-dependent integrals of motion $Q^\pm$ in the form
\begin{equation}\label{Q1}
Q^{\pm}=A^{\pm}\,e^{\mp\,i\,\alpha_{0}\,t}\ .
\end{equation}
Having in mind that the Hamiltonian $H$ is an integral of motion
whose value is the energy $E$ of the system, and the factorization
(\ref{haab}), the value of $Q^{\pm}$ is given by
\begin{equation}\label{q1}
q^\pm =\sqrt{E-\gamma_{0}}\, e^{\pm i\, \theta_0}
\end{equation}
where the energy is greater than the minimum of the potential,
$E>\gamma_0$. Then, taking into account (\ref{a1}), (\ref{Q1}) and
(\ref{q1}), we get the following equations
\begin{equation}\label{aqQ1}
\mp
i\,p+
\left(\frac{\alpha_{0}}{2}\, x+b_{0}\right)=
\sqrt{E-\gamma_{0}}\,e^{\pm
i\,(\theta_{0}+\alpha_{0}\,t)}.
\end{equation}
The phase trajectories ($x(t),p(t)$) can be found from these
equations and have the well known expressions
\begin{equation}\label{xt1}
\begin{array}{l}
\displaystyle
x(t)=
\displaystyle
\frac{2}{\alpha_{0}}\big(\sqrt{E-\gamma_{0}}\,\cos{(\theta_0+\alpha_{0}\,t)}-b_{0}\big)\\[2.ex]
p(t)=
\displaystyle
-\sqrt{E-\gamma_{0}}\,\sin{(\theta_0+\alpha_{0}\,t)}.
\end{array}
\end{equation}
The constant $b_0$ can be eliminated by a translation in the
$x$-axis, so hereafter in the next examples it will be set equal to
zero. The values of $q^\pm$ fix completely the initial conditions
$(x(0),p(0))$ of the motion.

\subsection{The Scarf potential}

Here we will consider  $(g\neq 0, \varphi=0)$, and the positive character
$H>0$ of the Hamiltonian for the bound states. Following the same
procedure as in the previous case, after some calculations we get the
Scarf potential
\begin{equation}\label{v2}
V(x)=\frac{\gamma_{0}}{\cos^{2}(\frac{\alpha_0}{2}x)}\,
,\qquad \gamma_0>0
\end{equation}
where $\gamma_{0}$ and $\alpha_{0}$ are constants. We also obtain
\begin{equation}\label{afp2}
\begin{array}{lll}
f(x)=\cos{(\frac{\alpha_{0}}{2}x)},\quad &
g(x)=\sin{(\frac{\alpha_{0}}{2}x)},\quad & \varphi(x)=0
\\[2.ex]
\alpha(H)=\alpha_{0}\,\sqrt{H},\quad &
\beta(H)=\alpha_{0}\,\sqrt{H},\quad &
\gamma(H)=\gamma_{0}\,  .
\end{array}
\end{equation}
Therefore, we can write the factor functions
\begin{equation}\label{a2}
A^{\pm}=\mp
i\,\cos{(\frac{\alpha_{0}}{2}x)}\,p+
\sqrt{H}\,\sin{(\frac{\alpha_{0}}{2}x)}
\end{equation}
the factorization,
\begin{equation}\label{fact2}
H = A^+ A^- +\gamma_0
\end{equation}
and the deformed algebra
\begin{equation}\label{ha2}
\{H,A^{\pm}\}=\pm i\,\alpha_{0}\,\sqrt{H}\,A^{\pm},\qquad
\{A^{+},A^{-}\}=- i\, \alpha_{0}\,\sqrt{H}.
\end{equation}
This algebra can be easily rewritten, defining $A_{0}\equiv\sqrt{H}$
as follows
\begin{equation}\label{nha2}
\{A_{0},A^{\pm}\}=\pm i\,\frac{\alpha_{0}}{2}\,A^{\pm},\qquad
\{A^{+},A^{-}\}=- i\, \alpha_{0}\,A_{0}
\end{equation}
which corresponds to the $su(1,1)$ Poisson algebra. We remark that
for the quantum system with Scarf potential the spectrum generating
algebra is also  $su(1,1)$ Lie algebra \cite{negro}.

Now, the time-dependent integrals of motion satisfying relation
(\ref{td}) in terms of $A^{\pm}$ are
\begin{equation}\label{q3}
Q^{\pm}=A^{\pm}\,e^{\mp\,i\,\alpha_{0}\,\sqrt{H}\,t}
\end{equation}
with the values $q^\pm = \sqrt{E -\gamma_{0}}\,e^{\pm i\,\theta_0}$.
As the energy must be greater than the minimum of the potential,
$E>\gamma_0$. Substituting (\ref{a2}) in (\ref{q3}) and taking into
account the values of the time-dependent integrals of motion, we get
\begin{equation}\label{xt2bp}
\begin{array}{l}
\displaystyle \mp
i\,p\,\cos{\frac{\alpha_{0}}{2}x}+\sqrt{E}\,\sin{\frac{\alpha_{0}}{2}x}=
\sqrt{E-\gamma_{0}}\,e^{\pm i(\theta_0+\alpha_{0}\, \sqrt{E}\,t)}.
\end{array}
\end{equation}
From these equations we have the following expressions
\begin{equation}\label{xt2bp1}
\begin{array}{l}
\displaystyle\sin{\frac{\alpha_{0}}{2}x}=\sqrt{\frac{E-\gamma_{0}}{E}}\,\cos{
(\theta_0+\alpha_{0}\,\sqrt{E}\,t)}\\[2.5ex]
\displaystyle p\,\cos{\frac{\alpha_{0}}{2}x}=-
\sqrt{E-\gamma_{0}}\,\sin{(\theta_0+\alpha_{0}\,\sqrt{E}\,t)}
\end{array}
\end{equation}
and the phase trajectories ($x(t),p(t)$) can be easily written in
the form:
\begin{equation}\label{xt2b}
\begin{array}{l}
\displaystyle
 x(t)=\frac{2}{\alpha_{0}}\,\arcsin(\sqrt{\frac{E-\gamma_{0}}{E}}
\cos{(\theta_0+\alpha_{0}\,\sqrt{E}\,t)})
\\[2.5ex]
\displaystyle
 p(t)=-\frac{\sqrt{E(E -\gamma_{0})}\,\sin{(\theta_0+\alpha_{0}\,
\sqrt{E}\,t)}}
{\sqrt{E -(E -\gamma_{0})\cos^{2}(\theta_0+\alpha_{0}\,\sqrt{E}\,t)}}
\ .
\end{array}
\end{equation}

\subsection{The P\"oschl-Teller potential}

In this section, we assume $(g\neq0, \varphi=0)$, and we choose the
negative sign $H<0$ of the Hamiltonian for the bound states.
Therefore, we must change the form of $A^\pm$ in (\ref{a}),
\begin{equation}\label{aaa}
A^{\pm}=\mp i f(x)\,p+\sqrt{-H}\,g(x)+\varphi(x)+\phi(H).
\end{equation}
The expressions (\ref{fg}), (\ref{hf}) and (\ref{b}) must also adapt
to this situation: replacing the root $\sqrt{H}$ by $\sqrt{-H}$ in
all these equations, and in particular expression (\ref{b}) takes
the form
\begin{equation}\label{3b}
\begin{array}{l}
\displaystyle
 \beta(H)=\sqrt{-H}\,\big[2
\,f(x)\,g'(x)-2\,f'(x)\,g(x)\big] - \frac{1}{\sqrt{-H}}\big[f(x)\,g(x)\,V'(x)+2\,f'(x)\,g(x)\,V(x)
\big]\\[2.ex]
\qquad
-4\,f'(x)\,\phi'(H)(H-V(x))+2\,f(x)\,\big[V'(x)\,\phi'(H)+\varphi'(x)\big].
\end{array}
\end{equation}
However, later we will also study the unbounded states with positive
energy. Following the same procedure as in the above cases and having in mind that $\phi(H)=0$, we arrive at the
P\"oschl-Teller potential
\begin{equation}\label{v3}
V(x)=-\frac{\gamma_{0}}{\cosh^{2}{(\frac{\alpha_{0}}{2}x)}}\,
, \qquad \gamma_0>0
\end{equation}
where $\gamma_{0}$ and $\alpha_{0}$ are constants. For this case, we
get
\begin{equation}\label{afp3}
\begin{array}{lll}
f(x)= \cosh{(\frac{\alpha_{0}}{2}x)},\quad &
g(x)= \sinh{(\frac{\alpha_{0}}{2}x)}, &\varphi(x) = 0
\\[2.ex]
\alpha(H)=\alpha_{0}\,\sqrt{-H},\quad &
\beta(H)= \alpha_{0}\,\sqrt{-H},\quad &
\gamma(H)=-\gamma_{0}\, .
\end{array}
\end{equation}
Then, we have
\begin{equation}\label{a3}
A^{\pm}=\mp
i\, \cosh{(\frac{\alpha_{0}}{2}x)}\,p+
\sqrt{-H}\,\sinh{(\frac{\alpha_{0}}{2}x)}
\end{equation}
the factorization
\begin{equation}\label{fac3}
H = A^+A^- -\gamma_0
\end{equation}
and the Poisson brackets
\begin{equation}\label{ha3}
\{H,A^{\pm}\}=\pm i\,\alpha_{0}\,\sqrt{-H}\,A^{\pm},\qquad
\{A^{+},A^{-}\}=-i\, \alpha_{0}\,\sqrt{-H}.
\end{equation}
This algebra can also be rewritten, defining $A_{0}\equiv-\sqrt{-H}$
as follows
\begin{equation}\label{nha2}
\{A_{0},A^{\pm}\}=\pm i\,\frac{\alpha_{0}}{2}\,A^{\pm},\qquad
\{A^{+},A^{-}\}=i\, \alpha_{0}\,A_{0}
\end{equation}
which corresponds to the $su(2)$ Poisson algebra. Here, we remark
that for the quantum P\"{o}shl-Teller potential the spectrum
generating algebra can also be identified with the $su(2)$ Lie algebra \cite{negro}.

Hence, the time-dependent integrals of motion  are
$Q^{\pm}=A^{\pm}\,e^{\mp\,i\,\alpha_{0}\,\sqrt{-H}\,t}$, with the
values $q^\pm=\sqrt{E +\gamma_{0}}\,e^{\pm i\, \theta_0}$. The
energy must be negative, but greater than the potential minimum,
$-\gamma_0 <E<0$. Replacing (\ref{a3}) in $Q^{\pm}$ and taking into
account the values $q^\pm$, we get
\begin{equation}\label{xt3bp}
\begin{array}{l}
\displaystyle \mp
i\,p\,\cosh{\frac{\alpha_{0}}{2}x}+\sqrt{-E}\,\sinh{\frac{\alpha_{0}}{2}x}=
\sqrt{E+\gamma_{0}}\,e^{\pm i(\theta_0+\alpha_{0}\, \sqrt{-E}\,t)} .
\end{array}
\end{equation}
Then, we have the following equations
\begin{equation}\label{xt3bp1}
\begin{array}{l}
\displaystyle\sinh{\frac{\alpha_{0}}{2}x}=\sqrt{-\frac{E-\gamma_{0}}{E}}\,\cos{
(\theta_0+\alpha_{0}\,\sqrt{-E}\,t)}\\[2.5ex]
\displaystyle p\,\cosh{\frac{\alpha_{0}}{2}x}=-
\sqrt{E-\gamma_{0}}\,\sin{(\theta_0+\alpha_{0}\,\sqrt{-E}\,t)}
\end{array}
\end{equation}
and from these equations ($x(t),p(t)$) can be found as
\begin{equation}\label{xt3}
\begin{array}{l}
\displaystyle x(t)=\frac{2}{\alpha_{0}}\,{\rm arcsinh}
(\sqrt{-\frac{E +\gamma_{0}}
{E}}\cos{(\theta_0+\alpha_{0}\,\sqrt{-E}\,t)})
\\[3ex]
\displaystyle
p(t)=-\frac{\sqrt{-E(E +\gamma_{0})}\,\sin{(\theta_0+\alpha_{0}\,\sqrt{-E}\,t)}}
{\sqrt{-E +(E +\gamma_{0})\cos^{2}(\theta_0+\alpha_{0}\,\sqrt{-E}\,t)}}.
\end{array}
\end{equation}

The potential (\ref{v3}) also allows for unbounded motions when
$E>0$. In this case we can write $\sqrt{-E}=i\,\sqrt{|E|}$ (or
$\sqrt{-H}=i\,\sqrt{H}$). Then, the factors $A^{\pm}$ are no longer
complex conjugate, but pure imaginary:
\begin{equation}\label{a33}
A^{\pm}=\mp i\, \cosh{(\frac{\alpha_{0}}{2}x)}\,p+i\,
\sqrt{H}\,\sinh{(\frac{\alpha_{0}}{2}x)}.
\end{equation}
The algebra of the Poisson brackets (\ref{ha3}) also changes for
this energy sector
\begin{equation}\label{ha33}
\{H,A^{\pm}\}=\mp\,\alpha_{0}\,\sqrt{H}\,A^{\pm},\qquad
\{A^{+},A^{-}\}= \alpha_{0}\,\sqrt{H}.
\end{equation}
But, defining $A_{0}\equiv -i\,\sqrt{H}$, this algebra has also the
form of an $su(2)$ Poisson algebra
\begin{equation}\label{ha331}
\{A_{0},A^{\pm}\}=\pm\,i\,\frac{\alpha_{0}}{2}\,A^{\pm},\qquad
\{A^{+},A^{-}\}= i\alpha_{0}\,A_{0}.
\end{equation}
Dealing with this potential in quantum mechanics, the bound states
are described by a finite-dimensional unitary representations of
$su(2)$ but the scattering states by infinite-dimensional
representations of the continuous series of $su(1,1)$
\cite{alhassid}. However, if we use the same kind of lowering and
raising operators, the scattering states belong to
infinite-dimensional non-unitary representations of $su(2)$. Here,
we have seen that in classical mechanics the bound motions are
described by the unitary $su(2)$ poisson algebra (in the sense that  the factor
functions $A^{\pm}$ are complex-conjugate) and the unbounded motions
by non-unitary $su(2)$ algebra (the factor functions $A^{\pm}$ are
pure imaginary)  by means of the Poisson brackets.

The integrals of motion $Q^{\pm}$ are imaginary and their
eigenvalues can be written in the form
 $q^{\pm}=\mp i\,\sqrt{E+\gamma_{0}}\,e^{\mp\,\theta_{0}}$.
After similar calculations, these integrals of motion have the explicit
expressions
\begin{equation}\label{xt3bp2}
\begin{array}{l}
\displaystyle \mp
p\,\cosh{\frac{\alpha_{0}}{2}x}+\sqrt{E}\,\sinh{\frac{\alpha_{0}}{2}x}=\mp
\sqrt{E+\gamma_{0}}\,e^{\mp (\theta_0+\alpha_{0}\, \sqrt{E}\,t)} .
\end{array}
\end{equation}
Therefore, we get
\begin{equation}\label{xt3bp3}
\begin{array}{l}
\displaystyle\sinh{\frac{\alpha_{0}}{2}x}=\sqrt{\frac{E+\gamma_{0}}{E}}\,\sinh{
(\theta_0+\alpha_{0}\,\sqrt{E}\,t)}\\[2.5ex]
\displaystyle p\,\cosh{\frac{\alpha_{0}}{2}x}=
\sqrt{E+\gamma_{0}}\,\cosh{(\theta_0+\alpha_{0}\,\sqrt{E}\,t)}.
\end{array}
\end{equation}
Finally, the unbounded non-periodic solutions are
\begin{equation}\label{xt31}
\begin{array}{l}
\displaystyle x(t)=\frac{2}{\alpha_{0}}\,{\rm arcsinh}
(\sqrt{\frac{E +\gamma_{0}}
{E }}\sinh{(\theta_0+\alpha_{0}\,\sqrt{E}\,t)})
\\[3ex]
\displaystyle
p(t)=
\frac{\sqrt{E(E +\gamma_{0})}\,\cosh{(\theta_0+\alpha_{0}\,\sqrt{E}\,t)}}
{\sqrt{E +(E +\gamma_{0})\sinh^{2}(\theta_0+\alpha_{0}\,\sqrt{E}\,t)}}.
\end{array}
\end{equation}

\begin{figure}[h]
  \centering
\includegraphics[width=0.4\textwidth]{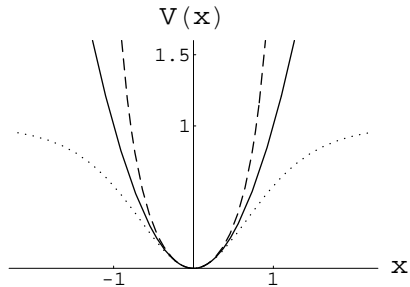}%
\caption{The oscillator (continuous line),
Scarf (dashed) and P\"oschl-Teller (dotted)
potentials.}
  \label{figura1}
\end{figure}

\begin{figure}[h]
  \centering
\includegraphics[width=0.75\textwidth]{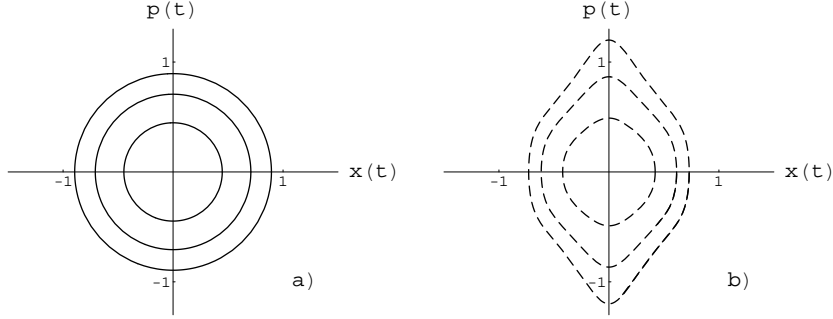}%
\caption{Some phase trajectories for the same three energies of
motions in the oscillator (a) and Scarf potentials (b). The exterior
trajectories of the latter have higher frequencies than the inner
ones.}
  \label{figura2}
\end{figure}

\begin{figure}[h]
\centering
\includegraphics[width=0.55\textwidth]{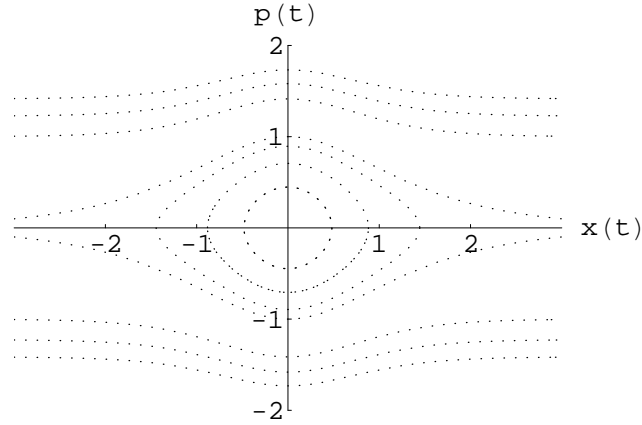}%
\caption{Now, it is shown the phase trajectories for
the same three energies
of periodic motion in the P\"oschl-Teller potential. The exterior
trajectories have slower frequencies than
the inner ones. Also it is shown some unbounded trajectories.}
\label{figura3}
\end{figure}

Remark the different dependence of the frequencies on energy for the
bound motion states of the above three systems. In the oscillator
case, it is a constant independent of $E$, for the Scarf potential
the frequency is an increasing function of $E$, while in the
P\"oschl-Teller potential the frequency is a decreasing function of
$E$. This corresponds to the values of the bracket (\ref{ha}) of the
Poisson algebra for each of these systems. The potentials and some
trajectories are plotted in Figures \ref{figura1}, \ref{figura2} and
\ref{figura3}. We also notice that the values of $\gamma_{0}$ in the
potentials (\ref{v2}) and (\ref{v3}) have been taken with suitable
signs giving rise to bound states.

\section{Singular potentials}

In this section we consider $\phi(H)\neq0$, to obtain other
classical systems. Here, we also propose some specific form of
$\phi(H)$ and consider $\gamma(H)$ different from constant so that
we will find the potentials of the previous section including a
singular term.

\subsection{The singular oscillator}

If we choose  $\phi(H)=\beta_{0}\,H$, and $(g(x)=0, \varphi(x)\neq
0)$, after similar calculations to the preceding cases, we get the
singular oscillator potential
\begin{equation}\label{vv}
V(x)=\frac{(\beta_{0}\,\alpha_{0}^{2}\,x-4\,b_{0})^2}{16\,\beta_{0}^{2}\,\alpha_{0}^{2}}+
\frac{4\alpha_0^2(\gamma_0+d_0^2)}{(\beta_{0}\,\alpha_{0}^{2}\,x-4\,b_{0})^2}-\frac{d_{0}}{\beta_{0}}
\end{equation}
where $\beta_{0}$, $\alpha_{0}$, $b_{0}$ and $d_{0}$ are constants.
Henceforth, we will choose the global constant $\beta_0=1$, the
potential origin $d_0=1/2$, and set the space origin $b_0 =0$, so
that we will handle the simplified expression
\begin{equation}\label{vvb}
V(x)=\frac{\alpha_{0}^{2}\,x^2}{16}+
\frac{4\gamma_0+1}{ \alpha_{0}^{2}\,x^2}-\frac{1}{2}
\ .
\end{equation}
For this choice, the unknown functions take the form
\begin{equation}\label{afpp1}
\begin{array}{ll}
f(x)=-\frac{1}{2 }\, \alpha_{0}^{2}\,x ,\quad
&\varphi(x)=-\frac{ \alpha_{0}^{2}\,x }{8 }+\frac12
\\[2.ex]
\beta(H)= 2\alpha_0 H -\alpha_{0},\quad
&
\gamma(H)=- H^{2}+ \gamma_{0}
\end{array}
\end{equation}
with $\alpha(H)=\alpha_{0}$.  Hence, the explicit expressions for
the factors are
\begin{equation}\label{aa1}
A^{\pm}=\pm
i\,\frac{ \alpha_{0} \,x }{2 }\,p-
\frac{ \alpha_{0}^{2}\,x^2}{8 }+ H+\frac12
\end{equation}
and the relation between the Hamiltonian and these functions
is given by
\begin{equation}\label{haab1}
H =
A^+A^{-}- H^{2}+ \gamma_{0}.
\end{equation}
Finally, we have the $su(1,1)$ Poisson algebra
\begin{equation}\label{haa1}
\{H,A^{\pm}\}=\pm i\,\alpha_{0}\,A^{\pm},\qquad
\{A^{+},A^{-}\}=- i\,\alpha_0\,(2 H -1) .
\end{equation}
In this case, the time-dependent integrals of motion $Q^\pm$ has the
form (\ref{Q1}), with $c(E)=\sqrt{ E^{2}+ E-\gamma_{0}}$. Then, the
phase trajectories ($x(t),p(t)$) for $E>\frac{- 1 +\sqrt{1 + 4
\gamma_{0}}}{2}$ are
\begin{equation}\label{xtt1}
\begin{array}{l}
\displaystyle
x(t)=\frac{2}{ \alpha_{0}}
\sqrt{ 2E+ 1 -2\, c(E)\,\cos{(\theta_0+\alpha_{0}\,t)}}
\\[2.5 ex]
\displaystyle
p(t)=\frac{ c(E)\,\sin{(\theta_0+\alpha_{0}\,t)}}
{\sqrt{ 2E+ 1 -2\, c(E)\,\cos(\theta_0+\alpha_{0}\,t)}}
\ .
\end{array}
\end{equation}
These expressions show that the frequency of the motion for the singular oscillator is a constant $\alpha_0$
independent of the energy $E$.

\subsection{The generalized Scarf potential ($H>0$)}

For the specific form
\begin{equation}
\phi(H)=\frac{\delta_{0}}{\sqrt{H}}+\beta_{0}\,\sqrt{H}
\end{equation}
and $\varphi(x)=0$, we have the generalized Scarf potential.
Here we will
assume $\beta_0\neq 0$, and by means of a global constant
of the factor functions we can set it equal to one: $\beta_0 =1$.
In this way we get the potential
\begin{equation}\label{v4}
V(x)=\frac{1}{4 }\left(\frac{2\, \delta_{0}
+
\gamma_{0}}{\sin^{2}{(\frac{\alpha_{0}}{4}x)}}-\frac{2 \,\delta_{0}
-
\gamma_{0}}{\cos^{2}{(\frac{\alpha_{0}}{4}x)}}\right)
\end{equation}
where $\gamma_{0}$, $\alpha_{0}$, and $\delta_{0}$ are
constants. We have also
\begin{equation}\label{fg4}
\begin{array}{lll}
\displaystyle f(x)=-2\, \cos{(\frac{\alpha_{0}}{4}x)}\,\sin{(\frac{\alpha_{0}}{4}x)},\
&
\displaystyle
g(x)=-2\, \sin^{2}{(\frac{\alpha_{0}}{4}x)},\ &
\varphi(x) = 0
\\[2.ex]
\displaystyle
\beta(H)= \alpha_{0}\,\sqrt{H}-\frac{\alpha_{0}\,
\delta_{0}^{2}}{H\,\sqrt{H}},\
&
\displaystyle
\gamma(H)=
\gamma_{0}- \frac{\delta_{0}^{2}}{H}, &
\alpha(H)=\alpha_{0}\,\sqrt{H}\ .
\end{array}
\end{equation}
Then, the two complex conjugate functions are
\begin{equation}\label{a4}
A^{\pm}=\mp
2\,i\, \cos{(\frac{\alpha_{0}}{4}x)}\,\sin{(\frac{\alpha_{0}}{4}x)}p
+
2\, \sqrt{H}\,\sin^{2}{(\frac{\alpha_{0}}{4}x)}-\frac{\delta_{0}}
{\sqrt{H}}- \sqrt{H}
\end{equation}
with the factorization relation
\begin{equation}\label{fac4}
H = A^+A^- +
\gamma_{0} - \frac{\delta_{0}^{2}}{H}
\end{equation}
and brackets
\begin{equation}\label{ha4}
\{H,A^{\pm}\}=\pm i\,\alpha_{0}\,\sqrt{H}\,A^{\pm},\qquad
\{A^{+},A^{-}\}=- i\,\alpha_{0}\,
\left( \sqrt{H}-\frac{ \delta_{0}^{2}}{H\,\sqrt{H}}\right).
\end{equation}
In this case, as well as in the remaining examples, the Poisson
algebra is not a Lie algebra, but a deformed algebra due to the
nonlinear function in the last bracket of (\ref{ha4}). We can write
the time-dependent integrals of motion in the form
$
Q^{\pm}=A^{\pm}\,e^{\mp\,i\,\alpha_{0}\,\sqrt{H}\,t}$, with the
value $q^{\pm}= c(E)\,e^{\pm\,i\,\theta_0}
$ 
where
\begin{equation}
c(E)=\sqrt{-\gamma_{0}+ E+\frac{\delta_{0}^{2}}{E}}.
\end{equation}
In order to have a positive potential with bound
states, if we assume $\gamma_0>0$, then the coefficient
$\delta_0$ must satisfy
\begin{equation}\label{con4}
-\gamma_0 < 2  \delta_0<\gamma_0\ .
\end{equation}
The energy of the bound states are given by the constraint $c(E)>0$,
that is, when $E$ is bigger than the minimum of $V(x)$,
\[E>\frac{\gamma_0 +\sqrt{\gamma_0^2 - 4 \delta_0^2}}
{2 } \ .
\]
Then, the trajectories ($x(t),p(t)$) are periodic for any allowed
value of $E$:
\begin{equation}\label{xt4}
\begin{array}{l}
\displaystyle
x(t)=\frac{2}{\alpha_{0}}\arccos{\big(-\frac{\delta_{0}+c(E)\,\sqrt{E}\,
\cos{(\theta_0+\alpha_{0}\,\sqrt{E}\,t)}}{ E}\big)}
\\[2.5ex]
\displaystyle
p(t)=-\frac{c(E)\,E\,\sin{(\theta_0+\alpha_{0}\,\sqrt{E}\,t)}}
{\sqrt{E^{2} - (\delta_{0}+c(E)\,\sqrt{E}\cos{(\theta_0+\alpha_{0}\,\sqrt{E}\,t)})^{2}}}.
\end{array}
\end{equation}

\subsection{The generalized P\"{o}schl-Teller potential ($H<0$)}

In this case, we consider $H<0$ and accordingly, we change the
sign of $H$ in the square roots of the function
$\phi(H)=\frac{\delta_{0}}{\sqrt{-H}}+\beta_{0}\,\sqrt{-H}$,
keeping $\varphi=0$. We also make use of expression (\ref{aaa}) of
$A^\pm$, and equation (\ref{3b}) of Section 3.3. In this way
we get the generalized P\"{o}schl-Teller potential
\begin{equation}\label{v5}
V(x)=\frac{1}{4 }
\left(\frac{2\, \delta_{0}+
\gamma_{0}}{\sinh^{2}{(\frac{\alpha_{0}}{4}x)}}
+\frac{2\, \delta_{0}-
\gamma_{0}}{\cosh^{2}{(\frac{\alpha_{0}}{4}x)}}\right)
\end{equation}
and the functions
\begin{equation}\label{fg5}
\begin{array}{lll}
\displaystyle f(x)=2\, \cosh{(\frac{\alpha_{0}}{4}x)}
\,\sinh{(\frac{\alpha_{0}}{4}x)},
&
\displaystyle
g(x)=2\, \sinh^{2}{(\frac{\alpha_{0}}{4}x)},
& \displaystyle
\varphi(x) = 0
\\[2.ex]
\displaystyle \beta(H)= - \alpha_{0}\,\sqrt{-H}-
\frac{\alpha_{0}\,\delta_{0}^{2}}{H\,\sqrt{-H}}, & \displaystyle
\gamma(H)=\gamma_{0}+2\,H + \frac{\delta_{0}^{2}}{H}, &
\displaystyle \alpha(H)=\alpha_{0}\,\sqrt{-H}
\end{array}
\end{equation}
where $\alpha_{0}$, $\delta_{0}$, $\gamma_{0}$ are constants, and we
have set $\beta_0=1$. Now, the factors $A^{\pm}$ are given by
\begin{equation}\label{a5}
A^{\pm}=\mp
2\,i\, \cosh{(\frac{\alpha_{0}}{4}x)}\,
\sinh{(\frac{\alpha_{0}}{4}x)}p
+
2\, \sqrt{-H}\,\sinh^{2}{(\frac{\alpha_{0}}{4}x)}+\frac{\delta_{0}}
{\sqrt{-H}}+ \sqrt{-H}
\end{equation}
with the factorization relation
\begin{equation}
-H = A^+A^- +
\gamma_{0} + \frac{\delta_{0}^{2}}{H}
\end{equation}
and the deformed Poisson algebra takes the form
\begin{equation}\label{ha5}
\{H,A^{\pm}\}=\pm i\,\alpha_{0}\,\sqrt{-H}\,A^{\pm},\qquad
\{A^{+},A^{-}\}=
i\,\alpha_{0}\,\left(\frac{ \delta_{0}^{2}}{H\,\sqrt{-H}}+
  \sqrt{-H}\right).
\end{equation}
The corresponding integrals of motion  in terms of $A^{\pm}$ are
$
Q^{\pm}=A^{\pm}\,e^{\mp\,i\,\alpha_{0}\,\sqrt{-H}\,t}$, with
$q^{\pm}=c(E)\,e^{\pm\,i\,\theta_0}
$,
where
\begin{equation}\label{c5}
c(E)=\sqrt{-\gamma_{0}- E -\frac{\delta_{0}^{2}}{E}}\ .
\end{equation}
If the potential has a positive singularity at $x=0$, in order to
allow for bound states, the coefficients in the potential, besides
(\ref{con4}), they must satisfy
$|2\, \delta_{0}+\gamma_{0}|<|2\, \delta_{0}-\gamma_{0}|$.
Then, the negative energy values of such states are given by the
positive character in the square root (\ref{c5}) of $c(E)$,
\begin{equation}\label{e5}
\frac{-\gamma_0 +\sqrt{\gamma_0^2 -
4 \delta_0^2}}{2 } < E< 0.
\end{equation}
Then, for these energies, the periodic trajectories ($x(t),p(t)$)
have the following form:
\begin{equation}\label{xt5}
\begin{array}{l}
\displaystyle x(t)=\frac{2}{\alpha_{0}} {\rm
arccosh}{\big(\frac{\delta_{0}-c(E)\,\sqrt{-E}
\cos{(\theta_0+\alpha_{0}\,\sqrt{-E}\,t)}}{ E}\big)}
\\[2.5ex]
\displaystyle
p(t)=-\frac{c(E)\,E\,\sin{(\theta_0+\alpha_{0}\,\sqrt{-E}\,t)}}
{\sqrt{(\delta_{0}-c(E)\,\sqrt{-E}\cos{(\theta_0+\alpha_{0}\,
\sqrt{-E}\,t)})^{2}- E^{2}}}\ .
\end{array}
\end{equation}

In order to get solutions for the unbounded motion, we choose $E>0$. Then,
the integrals of motion $Q^{\pm}$ are imaginary and the eigenvalues
can be written as $q^{\pm}= i\,c(E)\,e^{\mp\,\theta_{0}}$. In this
case the algebra (\ref{ha5}) changes its character. After similar
calculations, we have the unbounded phase trajectories ($x(t),p(t)$):
\begin{equation}\label{xt51}
\begin{array}{l}
\displaystyle x(t)=\frac{2}{\alpha_{0}} {\rm
arccosh}{\big(\frac{\delta_{0}+c(E)\,\sqrt{E}
\cosh{(\theta_0+\alpha_{0}\,\sqrt{E}\,t)}}{ E}\big)}
\\[2.5ex]
\displaystyle
p(t)=\frac{c(E)\,E\,\sinh{(\theta_0+\alpha_{0}\,\sqrt{E}\,t)}}
{\sqrt{(\delta_{0}+c(E)\,\sqrt{E}\cosh{(\theta_0+\alpha_{0}\,
\sqrt{E}\,t)})^{2}- E^{2}}}
\end{array}
\end{equation}
where
\begin{equation}\label{c51}
c(E)=\sqrt{\gamma_{0}+ E +\frac{\delta_{0}^{2}}{E}}.
\end{equation}

\begin{figure}[h]
\includegraphics[width=1.1\textwidth]{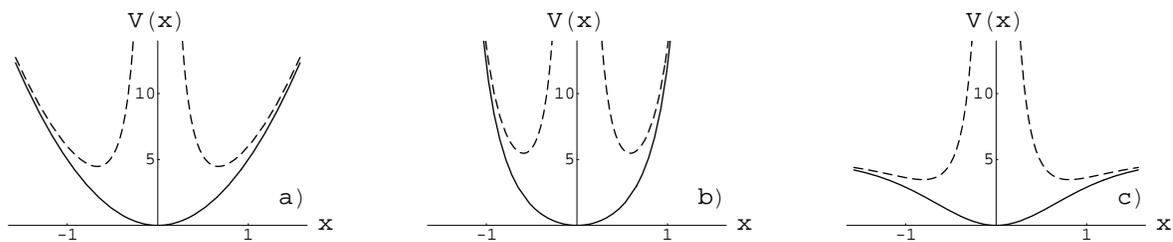}%
\caption{Plot of the oscillator (a), Scarf (b) and P\"oschl-Teller
(c) potentials (continuous lines) together with their singular
counterparts: the singular oscillator, generalized Scarf and
generalized P\"oschl-Teller potentials.}
  \label{figura4}
\end{figure}

\begin{figure}[h]
\centering
\includegraphics[width=0.5\textwidth]{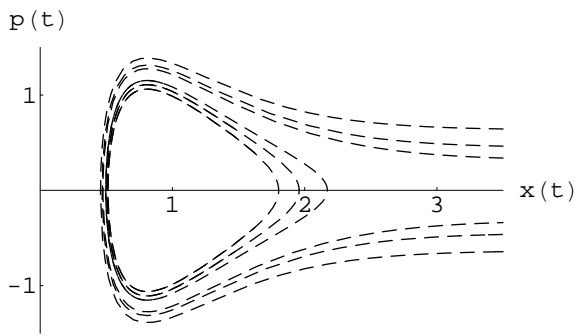}%
\caption{Plot of three closed and
three open trajectories in the phase space
for the generalized P\"oschl-Teller
potential.}
  \label{figura5}
\end{figure}

The figures of these singular potentials as well as some
trajectories in the phase space are depicted in Figures
\ref{figura4} and \ref{figura5}. Notice that the frequency
of the periodic motions in the above potentials is again
governed by $\alpha(E)$, and it has the same expression
as in the non-singular potentials of section 3.

\section{One dimensional Morse potentials}

In order to get the one dimensional Morse potential and other
related potentials, we will
consider in this section the
special form $\phi(H)=\frac{\delta_{0}}{\sqrt{H}}$  with
$\varphi=0$. This choice corresponds to the case excluded in the
previous section, when $\beta_{0}\to 0$.

\subsection{Hyperbolic case ($H<0$)}

For $\phi(H)=\frac{\delta_{0}}{\sqrt{-H}}$,
taking into account (\ref{aaa}) and (\ref{3b}),
we get the potential
\begin{equation}\label{v6}
V(x)=\frac{2\,\delta_{0}(C\,e^{\frac{\alpha_0}{2}x}+D\,
e^{-\frac{\alpha_0}{2}x}) -\gamma_{0}}
{(C\,e^{\frac{\alpha_0}{2}x}-D\,e^{-\frac{\alpha_0}{2}x})^{2}}
\end{equation}
where $\alpha_{0}$, $\delta_{0}$, $\gamma_{0}$ and $C,\,D$ are
constants. The functions $f(x)$, $g(x)$, $\beta(H)$ and $\gamma(H)$ have the following form
\begin{equation}\label{fg6}
\begin{array}{ll}
f(x)=C\, e^{\frac{\alpha_0}{2}x}-D\, e^{-\frac{\alpha_0}{2}x}, &
g(x)=C\,e^{\frac{\alpha_0}{2}x}+D\, e^{-\frac{\alpha_0}{2}x},\quad
\varphi(x) = 0
\\[2.ex]
\beta(H)=-\frac{\alpha_{0}\,\delta_{0}^{2}}{H\,\sqrt{-H}}-4\,C\,D\,\alpha_{0}\,\sqrt{-H},
&\gamma(H)=-\gamma_{0}+(4\,C\,D+1)\,H+\frac{\delta_{0}^{2}}{H}
\end{array}
\end{equation}
with $\alpha(H)=\alpha_{0}\,\sqrt{-H}$. Then, the poisson brackets
are
\begin{equation}\label{ha6}
\{H,A^{\pm}\}=\pm i\,\alpha_{0}\,\sqrt{-H}\,A^{\pm},\qquad
\{A^{+},A^{-}\}=
i\,(\frac{\alpha_{0}\,\delta_{0}^{2}}{H\,\sqrt{-H}}+4\,C\,D\,\alpha_{0}\,\sqrt{-H}).
\end{equation}
The functions $A^{\pm}$ can be written as
\begin{equation}\label{a8}
A^{\pm}=\mp
i\,(C\,e^{\frac{\alpha_0}{2}x}-D\,e^{-\frac{\alpha_0}{2}x})\,p+\sqrt{-H}
(C\,e^{\frac{\alpha_0}{2}x}+D\,e^{-\frac{\alpha_0}{2}x})+\frac{\delta_{0}}{\sqrt{-H}}
\end{equation}
and the time-dependent integrals of motion
\begin{equation}\label{q8b}
Q^{\pm}=A^{\pm}\,e^{\mp\,i\,\alpha_{0}\,\sqrt{-H}\,t},\qquad
q^{\pm}=c(E)\,e^{\pm\,i\,\theta_0}.
\end{equation}
We shall consider different cases according to the values of $C$ and
$D$, leading to regular potentials with bound states.

\subsubsection{$C=-D=1/2$.}
The option $C=-D$ is equivalent to any other with opposite signs of $C$ and $D$; it is
enough to consider an $x$-translation. By means of
a constant factor we can choose $C=-D=1/2$ then, the corresponding potential is known
as Scarf II (hyperbolic) \cite{cooper}
\begin{equation}\label{v9}
V(x)=2\,\delta_{0} \tanh{(\frac{\alpha_0}{2}x)}\,{\rm
sech}{(\frac{\alpha_0}{2}x)} - \gamma_{0}\,{\rm
sech}^{2}{(\frac{\alpha_0}{2}x)}.
\end{equation}
We have the following relations obtained for these particular values
of $C,D$:
\begin{equation}
H=A^+A^- -\gamma_{0}+\frac{\delta_{0}^{2}}{H}
\end{equation}
\begin{equation}\label{ha6}
\{H,A^{\pm}\}=\pm i\,\alpha_{0}\,\sqrt{-H}\,A^{\pm},\qquad
\{A^{+},A^{-}\}=
i\,\alpha_{0}\,(\frac{\delta_{0}^{2}}{H\,\sqrt{-H}}
-  \sqrt{-H}).
\end{equation}
For this case the functions $A^{\pm}$ read
\begin{equation}\label{a8}
A^{\pm}=\mp
i\,\cosh(\frac{\alpha_0}{2}x) \,p
+\sqrt{-H} \sinh(\frac{\alpha_0}{2}x) +\frac{\delta_{0}}{\sqrt{-H}}
\ .
\end{equation}
Then, ($x(t),p(t)$) have the following form:
\begin{equation}\label{xt8}
x(t)=\frac{2}{\alpha_{0}}{\rm
arcsinh}\big[\frac{\delta_0 -c(E)\,\sqrt{-E}\, \cos{
(\theta_0+\alpha_{0}\,\sqrt{-E}\,t)}}{E}\big]
\end{equation}
and
\begin{equation}\label{pt8}
p(t)=-\frac{c(E)\,E\, \sin{(\theta_0+\alpha_{0}\,\sqrt{-E}\,t)}}
{\sqrt{ E^2 +\big[\delta_{0}-c(E)\,\sqrt{-E}
\cos{(\theta_0+\alpha_{0}\,\sqrt{-E}\,t)\big]^{2}}}}
\end{equation}
where $c(E)=\sqrt{\gamma_{0}+ E-\frac{\delta_{0}^{2}}{E}}$. From
$c(E)$, we see that the range of the energy for bound states is
given by $(\gamma_0 -\sqrt{\gamma_0^2 + 4\delta_0^2})/2<E<0$. When
$E>0$, we have unbounded motion states with pure imaginary functions
$A^\pm$ and a different algebra from (\ref{ha6}), similar to that of
P\"oschl-Teller potential.

For the choice $C=D$, we get a trigonometric singular potential without bound states.

\subsubsection{$D=0$.}
In this case, by taking $D=0$, the corresponding potential and the functions
$f(x)$, $g(x)$, $\beta(H)$, and $\gamma(H)$ can be obtained from
(\ref{v6}) and (\ref{fg6}). We can also set $C=1$,
since $C$ is a global non-vanishing constant factor. The result
is the Morse potential
\begin{equation}
V(x)=2\,\delta_{0} \,e^{-\frac{\alpha_0}{2}x} -
\gamma_{0}\,e^{- \alpha_0 x} \ .
\end{equation}
Then, the complex conjugate functions are
\begin{equation}\label{a6}
A^{\pm}=\mp i\, e^{\frac{\alpha_0}{2}x}\,p+ \sqrt{-H}
e^{\frac{\alpha_0}{2}x}+\frac{\delta_{0}}{\sqrt{-H}}.
\end{equation}
Now, we can build the usual time-dependent integrals of motion
$Q^{\pm}$ with eigenvalues $q^{\pm}=c(E)\,e^{\pm\,i\,\theta_0}$
where $c(E)=\sqrt{-\frac{\delta_{0}^{2}}{E}+\gamma_{0}}$.
The periodic trajectories ($x(t),p(t)$) have the following form:
\begin{equation}\label{xt6}
x(t)=\frac{2}{\alpha_{0}}\log\big[\frac{\delta_{0}-c(E)\,\sqrt{-E}
\cos{(\theta_0+\alpha_{0}\,\sqrt{-E}\,t)}}{E}\big]
\end{equation}
and
\begin{equation}\label{pt6}
p(t)=-\frac{c(E)\,E\,\sin{(\theta_0+\alpha_{0}\,\sqrt{-E}\,t)}}{\delta_{0}-c(E)\,\sqrt{-E}
\cos{(\theta_0+\alpha_{0}\,\sqrt{-E}\,t)}}.
\end{equation}

In order to have periodic bounded motions we must take the
coefficients in the potential with the signs $\delta_0<0$
and $\gamma_0<0$. In this way the range of the energies is
$\delta_0^2/\gamma_0< E < 0$.

The case $C=0$ gives similar results, where the exponentials have
opposite signs than in the case just considered above.

\begin{figure}[h]
  \centering
\includegraphics[width=0.4\textwidth]{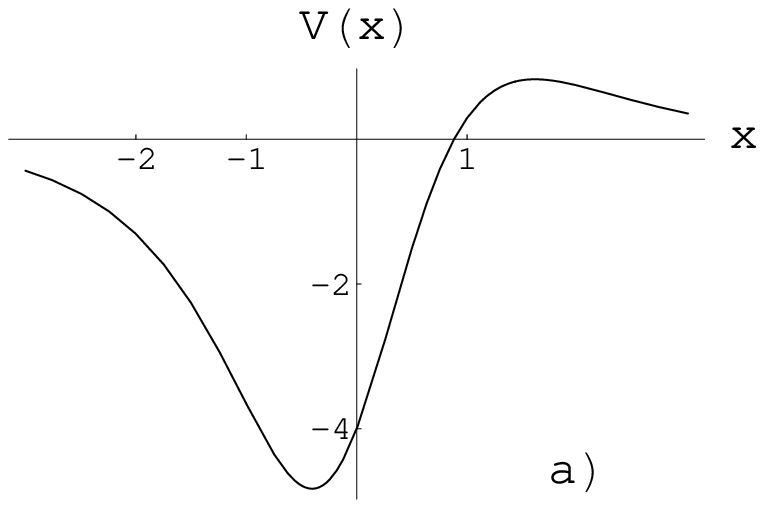}%
\hspace{1cm}%
  \includegraphics[width=0.4\textwidth]{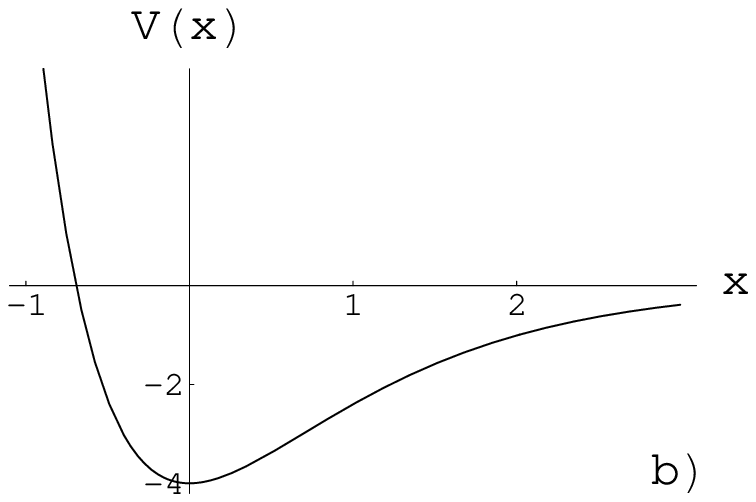}
\caption{Plot of the hyperbolic Scarf II potential
with the parameters $\gamma_0=4$, $\delta_0=2$ (a),
and the Morse potential with
$\gamma_0=-4$, $\delta_0=-4$, on the right (b).}
  \label{figuras6}
\end{figure}

\subsection{The trigonometric case ($H>0$)}
When we take $\phi(H)=\frac{\delta_{0}}{\sqrt{H}}$, we obtain the
potential
\begin{equation}\label{v10}
V(x)=-\frac{2\,\delta_{0}(C\,e^{i\,\frac{\alpha_0}{2}x}+D\,
e^{-i\,\frac{\alpha_0}{2}x}) +\gamma_{0}}
{(C\,e^{i\,\frac{\alpha_0}{2}x}-D\,e^{-i\,\frac{\alpha_0}{2}x})^{2}}
\end{equation}
where $\alpha_{0}$, $\delta_{0}$, $\gamma_{0}$ and $C,\,D$ are
constants. The involved functions have the following expressions
\begin{equation}\label{fg10}
\begin{array}{ll}
\displaystyle
f(x)=i(C\, e^{i\,\frac{\alpha_0}{2}x}-D\,
e^{-i\,\frac{\alpha_0}{2}x}),
& \displaystyle
g(x)=C\,e^{i\,\frac{\alpha_0}{2}x}+D\,
e^{-i\,\frac{\alpha_0}{2}x},\quad \varphi(x) = 0
\\[2.ex]
\displaystyle
\beta(H)=-\frac{\alpha_{0}\,\delta_{0}^{2}}{H\,\sqrt{-H}}+4\,C\,D\,\alpha_{0}\,\sqrt{H},
& \displaystyle
\gamma(H)=\gamma_{0}+(1-4\,C\,D)\,H-\frac{\delta_{0}^{2}}{H}
\end{array}
\end{equation}
with $\alpha(H)=\alpha_{0}\,\sqrt{H}$. Then, the associated Poisson brackets are
\begin{equation}\label{ha10}
\{H,A^{\pm}\}=\pm i\,\alpha_{0}\,\sqrt{H}\,A^{\pm},\qquad
\{A^{+},A^{-}\}=
i\,(\frac{\alpha_{0}\,\delta_{0}^{2}}{H\,\sqrt{H}}-4\,C\,D\,\alpha_{0}\,\sqrt{H}).
\end{equation}
For this case $A^{\pm}$ take the form
\begin{equation}\label{a10}
A^{\pm}=\pm
(C\,e^{i\,\frac{\alpha_0}{2}x}-D\,e^{-i\,\frac{\alpha_0}{2}x})\,p+\sqrt{H}
(C\,e^{i\,\frac{\alpha_0}{2}x}+D\,e^{-i\,\frac{\alpha_0}{2}x})+\frac{\delta_{0}}{\sqrt{H}}
\end{equation}
and the time-dependent integrals of motion are given by
$
Q^{\pm}=A^{\pm}\,e^{\mp\,i\,\alpha_{0}\,\sqrt{H}\,t}$,
with 
$q^{\pm}=c(E)\,e^{\pm\,i\,\theta_0}
$.

When we take $D=-C=i/2$, we obtain the potential called Scarf I
(trigonometric) \cite{cooper}
\begin{equation}\label{v11}
V(x)= 2 \delta_{0} \, \tan{(\frac{\alpha_0}{2}x)}\,{\rm
sec}{(\frac{\alpha_0}{2}x)} +
 \gamma_{0}\,{\rm sec}{(\frac{\alpha_0}{2}x)}^{2}
\end{equation}
together with
\begin{equation}\label{fg10b}
\begin{array}{lll}
f(x)=\cos(\frac{\alpha_0}{2}x), &
g(x)=\sin(\frac{\alpha_0}{2}x),\quad &\varphi(x) = 0
\\[2.ex]
\beta(H)=-\frac{\alpha_{0}\,\delta_{0}^{2}}{H\,\sqrt{-H}}
+ \alpha_{0}\,\sqrt{H},
&\gamma(H)=\gamma_{0} - \frac{\delta_{0}^{2}}{H},
&\alpha(H)=\alpha_{0}\,\sqrt{H}\ .
\end{array}
\end{equation}
Then, the Poisson brackets are
\begin{equation}\label{ha10b}
\{H,A^{\pm}\}=\pm i\,\alpha_{0}\,\sqrt{H}\,A^{\pm},\qquad
\{A^{+},A^{-}\}=
i\,\alpha_{0}\,(\frac{\delta_{0}^{2}}{H\,\sqrt{H}}-
 \sqrt{H}).
\end{equation}
Now the functions $A^{\pm}$ take the form
\begin{equation}\label{a10b}
A^{\pm}=\mp i
\cos(\frac{\alpha_0}{2}x)\,p+\sqrt{H}
\sin( \frac{\alpha_0}{2}x)+\frac{\delta_{0}}{\sqrt{H}}
\end{equation}
and
\begin{equation}
c(E) = \sqrt{E + \frac{\delta_0^2}{E} - \gamma_0}\ .
\end{equation}
From this expression we find that bound motion states exist when $
\gamma_0 > 2\delta_0 $. Under this condition, the allowed energies
are $E> (\gamma_0 + \sqrt{\gamma_0^2-4\delta_0^2})/2$. The
corresponding motions can be found without difficulty as in the
Scarf potential.

\begin{figure}[h]
  \centering
\includegraphics[width=0.4\textwidth]{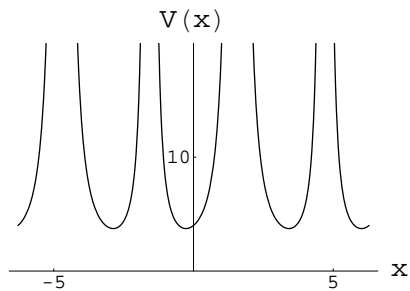}%
\caption{Plot of the trigonometric Scarf I potential for the values:
$\delta_0 = 1$, $\gamma_0=4$.}
  \label{figuras6}
\end{figure}

\section{Conclusions}

In this work we have studied a whole class of one-dimensional
classical systems characterized by an underlying Poisson algebra
which in general is  a deformed Lie algebra. Here, the Poisson
algebra is not made up of (time-independent) integrals of motion, as
it is the usual case \cite{bonatsos,daskaloyannis, daskaloyannis2},
but it includes functions, $A^\pm$, directly related with
time-dependent integrals of motion.

We have obtained some systems like the Scarf, P\"{o}schl-Teller, Morse
etc.~ which are clearly the classical analogues of the
one-dimensional quantum systems that can be solved by means of the
factorization method. However, the factorizations here employed are
not exactly the factorizations of the corresponding quantum cases
\cite{cooper}. Instead, they are more related with the so called
spectrum generating algebras in quantum mechanics
\cite{alhassid,raab,negro}, as can be seen replacing Poisson
brackets by commutators: $\{\cdot,\cdot\}\to -i[\cdot,\cdot]$.
Therefore, we have shown that these algebras can also be useful in
classical mechanics to compute time-dependent integrals of motion of
Bohlin type that give us the solutions of the motion in the phase
space
\cite{gettys,ray,pilar}.

The algebraic structures here obtained for some classical systems
correspond to well known ones for the analog quantum mechanical
systems. Hence, this correspondence is very important to describe
the coherent states of such quantum systems: for instance, the
expected values of $x$ and $p$ are adequate for the harmonic
oscillator, but in the case of the Scarf potential we should
consider the expected values of $\sin{x}$ and $p\,\cos{x}$, which
are the components of the functions $A^{\pm}$, as we see from
(\ref{xt1}) and (\ref{xt2bp1}), respectively \cite{negro,nieto}.

\section*{Acknowledgements}
This work is supported by the Spanish MEC (FIS2005-03989), AECI-MAEC (\c{S}~K grant 0000169684) and Junta de Castilla y Le\'on (Excellence project VA013C05). \c{S}~K  acknowledges Department of Physics, Ankara
University and also the warm hospitality at Department of
Theoretical Physics, University of Valladolid, Spain.


\end{document}